\documentclass[onecolumn,showpacs,twoside,pra]{revtex4}
\usepackage{mathrsfs}
\usepackage{amssymb, amsbsy, amsmath, latexsym, dsfont, array, layout, graphicx,mathrsfs,color}

\newcommand{\ket}[1]{\left|{#1}\right\rangle}
\newcommand{\bra}[1]{\left\langle{#1}\right|}

\begin{document}

\title{Spin Squeezing Property of Weighted Graph States}
\author{Peng Xue}
\affiliation{Department of Physics, Southeast University, Nanjing
211189, China}
\date{\today}

\begin{abstract}
We study the spin squeezing property of weighted graph states, which
can be used to improve the sensitivity in interferometry.
Decoherence reduces the spin squeezing property but the result
remains superior over other reference schemes with GHZ-type
maximally entangled states and product states. We study the time
evolution of spin squeezing of weighted graph states coupled to
different decoherence channels. Based on the analysis, the spin
squeezing of the weighted graph states is robust in the presence of
decoherence and the decoherence limit in the improvement of the
interferometric sensitivity is still achievable.
\end{abstract}


\maketitle
\section{Introduction}
Quantum correlations are important resources for nonclassical
phenomena and quantum information processes. Such as spin squeezing
and GHZ-type maximal entanglement have also found many promising
applications such as achieving interferometric
\cite{C81,Y86,KU91,UK01} and spectroscopic \cite{WBIMH92}
sensitivities beyond the standard quantum noise limit. Weighted
graph states are quantum correlated states \cite{CL09} with
fluctuations reduced in one of the collective spin components and
have possible applications in atomic interferometers and
high-precision atomic clocks. Whereas entanglement is based on the
superposition principle combined with the Hilbert space structure,
while squeezing is originated from another fundamental principle of
quantum mechanics---the uncertainty principle. Therefore, it is
interesting and important to study their relationship and different
applications. It is found that spin squeezing and entanglement are
closely related and spin squeezing implies
entanglement~\cite{MWSN11,WS03}. In this paper we will study their
different robustness to the quantum noise and decoherence.

To evaluate the potential application of quantum correlations such
as spin squeezing and entanglement, it is therefore essential to
include a realistic description of noise in experiments of
interests. In this paper, we analyze the effect of realistic
decoherence processes and noise sources on shot-noise phase
sensitivity $\delta \varphi$ in estimation of a collective rotation
angle $\varphi$.

In the absence of decoherence, the squeezed uncertainty of a spin
component directly improves the sensitivity in interferometry
\cite{KU91}, enabling the improvement factors up to $N^{1/3}$ for
one-axis twisting and $N^{1/2}$ for two-axis twisting with $N$
particles \cite{KU93}. GHZ-type maximally entangled states can also
improve interferometric sensitivity by a factor of $N^{1/2}$
treating $N$ particles as a single quantum object. The phase
evolution of an object consisting of $N$ particles is $N$ times as
fast as that of a particle itself, or an equivalently deBroglie
wavelength is shortened by a factor of $N$, giving the standard
quantum limit for an object $\sim1$ \cite{JBCY95}.

In the presence of decoherence, a GHZ type maximally entangled state
with zero spin squeezing parameter does not provide higher
resolution compared to product states\cite{UK01,HM97,ASL04,PS09}.
Whereas a partially entangled state with a high symmetry such as a
weighted graph state gives an improved sensitivity. Thus in the
presence of decoherence the spin squeezing results the improvement
the sensitivity in interferometry. To find weighted graph states
whose spin squeezing property is robust to decoherence
\cite{WMLSN10,S11} is the aim of this paper. Graph states proved
very robust against many sources of noise \cite{BR01,HDB04,RJ09}. In
this paper we will study a special kind of graph states--weighted
graph states and combine the robustness of the spin squeezing
property of these states with a scheme capable of improving
interferometric sensitivity.

A weighted graph state is a multipartite entangled state consisting
of a set of vertices $j$ connected to each other by edges taking the
form of controlled phase gate~\cite{DHHLB05,HCDB07}. The qubits are
prepared in $\ket{+}_j$, where
$\ket{\pm}_j=\left(\ket{\downarrow}\pm\ket{\uparrow}\right)_j/\sqrt{2}$
and $\left\{\ket{\downarrow}_j,\ket{\uparrow}_j\right\}$ is the
single-qubit computational basis. The weights
$\alpha_{jk}\in\left[0,2\pi\right]$ correspond to a controlled phase
operation applied between the two vertices $j$ and $k$. Weighted
graph states are a $O\left(N^2\right)$ parameter family of $N$-spin
state, correspond to weighted graphs which are independent of the
geometry and adaptable to arbitrary geometries and spatial
dimensions~\cite{APDVB06}.

This paper is organized as follows. Firstly we introduce the
definitions of spin squeezing parameters in Sec.~II. In Sec.~III,
the generation of weighted graph states and the spin squeezing
property of weighted graph states are introduced. The evolution of
the spin squeezing of the weighted graph states coupled to three
kinds of decoherence channels is shown in Sec.~IV and then we
analyze the robustness of the spin squeezing of weighted graph
states against decoherence. In Sec.~V, we discuss the application of
the spin squeezing property of weighted graph states---to improve
the frequency standard in the presence of decoherence. Finally we
briefly summarize the paper in Sec.~VI.

\section{Spin Squeezing Definitions}
The definition of spin-squeezing is not unique, and in the section
we review two most widely studied squeezing parameters proposed by
Kitagawa and Ueda \cite{KU91} and by Wineland et al. \cite{WBIMH92}.

We consider an ensemble of $N$ two-level particles with lower
(upper) state $\ket{\downarrow}$ ($\ket{\uparrow}$). Adopting the
nomenclature of spin-$1/2$ particles, we introduce the total angular
momentum (i.e., Bloch vector)
\begin{equation}
\vec{J}=\sum_{j=1}^N\vec{S}_j,
\end{equation}
where
\begin{equation}
\vec{S}_z^j=\frac{1}{2}\hat{\sigma}_z^j=\frac{1}{2}\left(\ket{\uparrow}_j\bra{\uparrow}-\ket{\downarrow}_j\bra{\downarrow}\right).
\end{equation}
At this point, it is convenient to introduce the following two kinds
of spin squeezing parameters \cite{WBIMH92,AP94}
\begin{align}
&\xi_1^2=\frac{4\big(\Delta
J_{\vec{n}_\bot}\big)_\text{min}^2}{N},\\ \nonumber
&\xi_2^2=\frac{N^2}{4\langle\vec{J}\rangle^2}\xi_1^2.
\end{align}
Here, the minimization in the first equation is over all directions
denoted by $\vec{n}_\bot$, perpendicular to the mean spin direction
$\vec{n}=\langle\vec{J}\rangle/|\langle\vec{J}\rangle|$. If
$\xi_2^2<1$ is satisfied, the spin squeezing occurs and the
$N$-qubit state is entangled.

It is known that the spin squeezing manifested by $\xi_2^2<1$ leads
to improvement in the frequency measurement \cite{HM97,UK01}.
Generally speaking, an interferometer is quantum mechanically
described as a collective, linear, rotation of input state by an
angle $\varphi$:
\begin{equation}
\rho_\text{out}\left(\varphi\right)=e^{\text{i}\varphi\vec{J}_{\vec{n}}}
\rho_\text{in}e^{-\text{i}\varphi\vec{J}_{\vec{n}}}.
\end{equation}
The goal is to estimate $\varphi$ with a sensitivity overcoming the
standard quantum limit or shot-noise limit
$\delta\varphi=1/\sqrt{N}$ \cite{PS09}.

Spin squeezed states are multiqubit entangled states showing
pairwise entanglement and reduced uncertainty in the collective spin
moment in one direction. This reduction in the measurement
uncertainty, achieved without violating the minimum uncertainty
principle by a redistribution of the quantum fluctuations between
noncommuting variables can be exploited to perform metrology beyond
the Heisenberg limit. The degree of squeezing of a spin ensemble is
evaluated by the squeezing parameter $\xi_2^2$. When spin squeezing
is instead used in the context of quantum limited metrology, the
squeezing parameter should measure the improvement in
signal-to-noise ratio for the measured quantity $\varphi$ and the
phase sensitivity is
\begin{equation}
\delta\varphi=\frac{\xi_2}{\sqrt{N}}.
\end{equation}
This definition is associated to Ramsey-type experiments
\cite{Ramsey}, in which an external magnetic field is measured via
the detection of the accumulated phase $\varphi$ due to the Zeeman
interaction and the phase uncertainty for a product state is
$\sim\sqrt{N}$, and for a GHZ-type maximally entangled state is
$\sim N$ in the absence of decoherence. If $\xi_2<1$, $\delta
\varphi$ beats the shot-noise limit. In the limit of large spin
numbers, using the one- and two-axis squeezing operator, the optimal
squeezing parameters are $\xi_1^2\sim3^{2/3}/(2N^{2/3})$ \cite{KU93}
and $\xi_2^2\sim(1+2\sqrt{3})/(2N)$ \cite{AL02,SGDM03},
respectively.

\section{Spin Squeezing Property of Weighted Graph States}
We introduce so-called weighted graph states in the section. We
begin by defining a graph, $G=\left(V,E\right)$, as a finite
non-empty point set $V$ with a collective $E\subset V$ of unordered
pairs of points in $V$. And $V$ is the collection of edges. Graphs
such as these can be used to describe a family of quantum states in
the following manner. The initial states are prepared in
$\ket{+}^{\otimes N}$. Between each pair of qubits connected by an
edge in the associated graph, we apply a unitary
operation---controlled phase operation. Weighted graph states
naturally arise when spin system interact via an Ising-type
interaction.

We consider $N$ spin-$1/2$ system with pairwise interactions
described by an Ising-type Hamiltonian
\begin{equation}
\hat{H}=\sum_{j,k}\frac{1}{4}f(j,k)\left(\mathbb{I}-\hat{\sigma}_z^j\right)
\otimes\left(\mathbb{I}-\hat{\sigma}_z^k\right).
\end{equation}
The weighted graph state \cite{BR01,HDB04} is obtained by the
evolution of the above Hamiltonian
\begin{align}
\ket{\psi_t}=\exp\left(-\text{i}\hat{H}t\right)\ket{+}^{\otimes N}
\propto\prod_{j,k}\exp\left[-\frac{1}{4}\text{i}f(j,k)t\hat{\sigma}_z^j\hat{\sigma}_z^k\right]\ket{+}^{\otimes
N}.
\end{align}
If we choose the evolution time to satisfy $f(j,k)t=m\pi$ with $m$
an integer, the state $\ket{\psi_t}$ is a product state. If
$f(j,k)t=\left(2m+1\right)\pi/2$, $\ket{\psi_t}$ becomes a graph
state. And for $0<f(j,k)t<\pi/2$, $\ket{\psi_t}$ is a weighted graph
state with spin squeezing parameter $\xi_2<1$.

\begin{figure}
   \includegraphics[width=8.5cm]{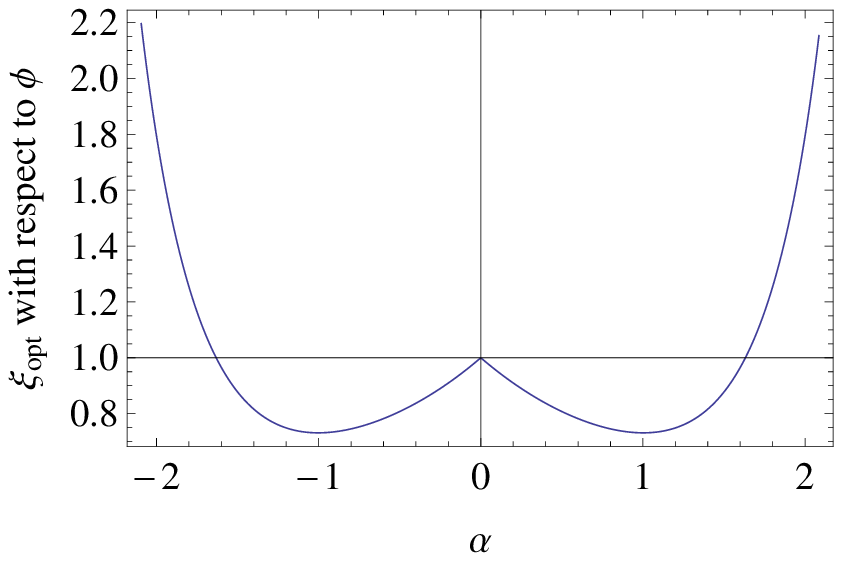}
   \caption{ Simulation of spin squeezing parameter for weighted cluster
   state $\xi_2$ minimized with respect to $\phi$ as a function of $\alpha$.}
   \label{figure1}
\end{figure}

\begin{figure}
   \includegraphics[width=8.5cm]{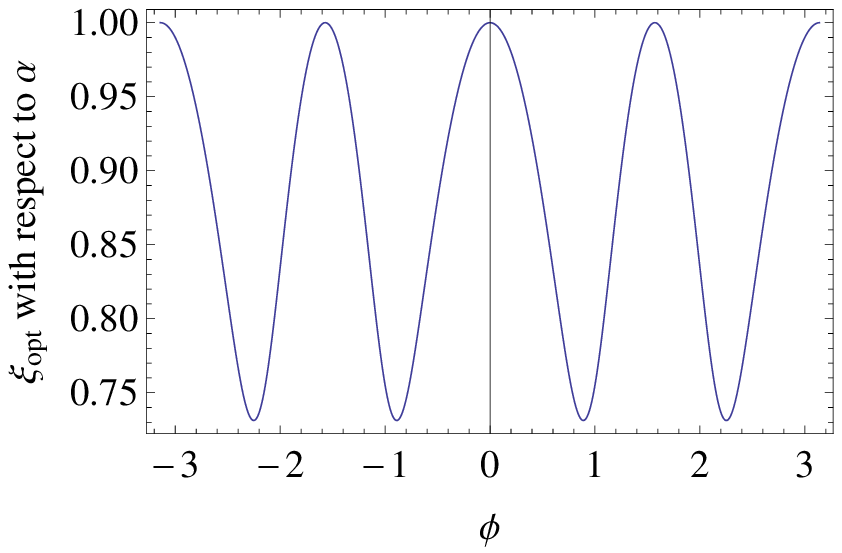}
   \caption{ Simulation of spin squeezing parameter for weighted cluster
   state $\xi_2$ minimized with respect to $\alpha$ as a function of $\phi$.}
   \label{figure2}
\end{figure}

In this paper we only consider two kinds of simplest weighted graph
states---weighted cluster states and weighted fully-connected graph
states. For the weighted cluster state we have the form
\begin{equation}
\ket{\psi_t}=\prod_{j=1}^{N-1}\exp\left[-\frac{\text{i}}{4}
f_jt\left(\mathbb{I}-\hat{\sigma}_z^j\right)\otimes\left(\mathbb{I}-\hat{\sigma}_z^{j+1}\right)\right]\ket{+}^{\otimes
N}.
\end{equation}
If we make our case further simpler by assuming $f_j\equiv f$ for
each qubit $j$ of the weighted cluster state and $ft\equiv\alpha$,
the spin squeezing parameter takes this form
\begin{equation}
\xi_2^2=\frac{1+\cos^2\phi\sin^2\alpha/2+\sin2\phi\sin\alpha}{\cos^4\alpha/2}
\end{equation}
with the mean spin direction
\begin{equation}
\vec{n}=\left(\cos\alpha,\sin(-\alpha),0\right),
\end{equation}
the orthogonal direction
\begin{equation}
\vec{n}_\perp=\left(-\cos\phi\sin(-\alpha),\cos\phi\cos\alpha,\sin\phi\right),
\end{equation}
and $\phi\in\left[-\pi,\pi\right]$. The spin squeezing parameter is
independent of the number of the qubits $N$ and can be optimized to
be $\xi_\text{min}=0.7312$ with respect to $\alpha$ and $\phi$ shown
in Figs.~1 and 2.

\begin{figure}
   \includegraphics[width=8.5cm]{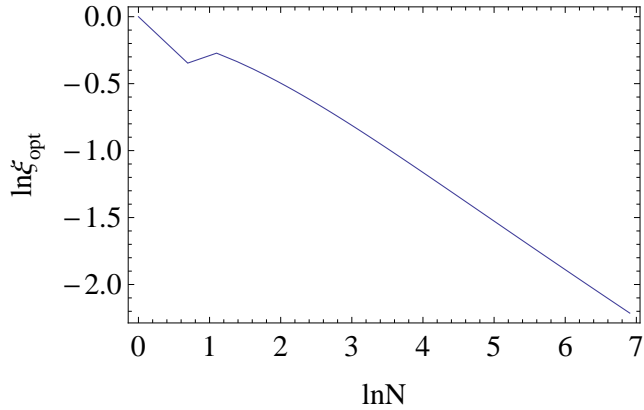}
   \caption{Ln-ln plot of the spin squeezing parameter for weighted fully-
   connected graph state $\xi_2$ v.s. the number of the qubits $N$
   (up to $10^3$) optimized
   with respect to $\alpha$.}
   \label{figure3}
\end{figure}

For the weighted fully-connected graph state, we have the form
\begin{equation}
\ket{\psi_t}=\prod_{j\neq
k}\exp\left[-\frac{\text{i}}{4}\alpha\left(\mathbb{I}-\hat{\sigma}_z^j\right)\otimes
\left(\mathbb{I}-\hat{\sigma}_z^k\right)\right]\ket{+}^{\otimes N}.
\end{equation}
The spin squeezing parameter of the weighted fully-connected graph
state with the same weight $\alpha$ takes this form
\begin{equation}
\xi_2^2=\frac{1-\left(N-1\right)\left[\sqrt{A^2+B^2}-A\right]/4}{\cos^{2N-2}\alpha},
\end{equation}
where
\begin{equation}
A=1-\cos^{N-2}\left(2\alpha\right), B=4\sin\alpha\cos^{N-2}\alpha.
\end{equation}
The mean spin direction for the weighted fully-connected graph state
is
\begin{equation}
\vec{n}=\left(\cos\left(N\alpha\right),\sin\left(-N\alpha\right),0\right),
\end{equation}
and the orthogonal direction is then
\begin{equation}
\vec{n}_{\perp}=\left(-\cos\phi\sin\left(-N\alpha\right),\cos\phi\cos\left(N\alpha\right)
,\sin\phi\right).
\end{equation}
The minimum spin squeezing parameter with respect to $\alpha$ is
obtained $\xi_2\propto 1/N^{1/3}$ shown in Fig.~3.

\section{Evolution of Spin Squeezing of Weighted Graph States in The
Presence of Decoherence}

The decoherence effects are described by three types of decoherence
channels: the amplitude damping channel, the dephasing channel, and
the depolarizing channel. In general, decoherence process can be
described by these three typical channels.

We consider a single qubit coupled to an environment which is
described by a thermal reservoir. The evolution of this qubit is
governed by a general master equation of Lindblad form
\begin{equation}
\frac{\partial \chi}{\partial
t}=-i\left[\hat{H}_r,\chi\right]+\mathcal{L}\chi,
\end{equation}
where the reference system is the standard system used in phase
sensitivity and it is defined as
\begin{equation}
\hat{H}_\text{r}=\frac{\Delta}{2}\sum_{j=1}^N\hat{\sigma}_z^j.
\end{equation} Whereas, the incoherent processes are described by
the superoperator $\mathcal{L}$:
\begin{align}
\mathcal{L}\chi=&-\frac{b}{2}(1-s)\left[\hat{\sigma}_+\hat{\sigma}_-\chi+
\chi\hat{\sigma}_+\hat{\sigma}_--2\hat{\sigma}_-\chi\hat{\sigma}_+\right]
-\frac{b}{2}s\left[\hat{\sigma}_-\hat{\sigma}_+\chi+\chi\hat{\sigma}_-\hat{\sigma}_+
-2\hat{\sigma}_+\chi\hat{\sigma}_-\right]\nonumber\\&-\frac{2c-b}{8}
\left[2\chi-2\hat{\sigma}_z\chi\hat{\sigma}_z\right],
\end{align}
with $\hat{\sigma}_\pm=\left(\hat{\sigma}_x\pm
i\hat{\sigma}_y\right)/2$. For an arbitrary $s$, $b=0$ and
$c=\gamma$, the qubit is coupled to a dephasing channel. For $s=1/2$
and $b=c=\gamma$, the qubit is coupled to a depolarizing channel.
Whereas, for $s=1$ and $b=2c=\gamma$, that is coupled to a decay
channel (pure damping).

Equivalently, one can use the resulting completely positive map
$\mathcal{E}$ with $\chi'=\mathcal{E}\chi$ as follows:
\begin{equation}
\mathcal{E}\chi=\sum_{j=0}^3 p_j(t)\hat{\sigma}_j\chi\hat{\sigma}_j.
\end{equation}
with $\chi$ a density matrix for a single-qubit state and
$\sum_{j=0}^3p_j(t)=1$. These noise channels are of particles
interest in quantum information theory, especially in the study of
fault-tolerance of quantum computation. This class contains for
example: (i) for $p_0=(1+3\text{e}^{-\gamma t})/4$ and
$p_1=p_2=p_3=(1-\text{e}^{-\gamma t})/4$ the above depolarizing
channel; (ii) for $p_0=(1+\text{e}^{-\gamma t})/2$, $p_1=p_2=0$ and
$p_3=(1-\text{e}^{-\gamma t})/2$ the above dephasing channel.
Finally, the decay channel is obtained:
\begin{equation}
\mathcal{E}\chi=E_0\chi E_0^\dagger+E_1\chi E_2^\dagger,
\end{equation}
with the Kraus operators $E_0=\begin{pmatrix}
            1 & 0\\
            0 & e^{-\gamma t/2} \\
        \end{pmatrix}$ and $E_1=\begin{pmatrix}
            0 & \sqrt{1-e^{-\gamma t}}\\
            0 & 0 \\
        \end{pmatrix}$~\cite{HDB04}.

For a system consisting of $N$ two-level particles, we would be
interested in the effect of decoherence on the spin squeezing
properties of this system. We consider as a decoherence model
individual coupling of each of the qubits to a thermal bath, where
the evolution of the $k$th qubit is described by the map
$\mathcal{E}_k$ with Pauli operators $\hat{\sigma}_j$ ($j=0,1,2,3$)
acting on qubit $k$. We would be interested in the evolution of a
given weighted graph state $\psi$ of $N$ qubit under this
decoherence model. That is, the initial state $\psi$ suffers from
decoherence and evolves in time to a mixed state $\rho(t)$ given by
\begin{equation}
\rho(t)=\mathcal{E}_1\mathcal{E}_2...\mathcal{E}_N\ket{\psi_t}\bra{\psi_t}.
\end{equation}

Decoherence destroys entanglement and thus the improvement in
sensitivity via entangled states (especially via GHZ-type maximally
entangled states) is reduced in practice. Whereas, weighted graph
states with spin squeezing robust to decoherence can still provide
an improvement in sensitivity compared to GHZ-type maximally
entangled states and product states.

\begin{figure}
   \includegraphics[width=8.5cm]{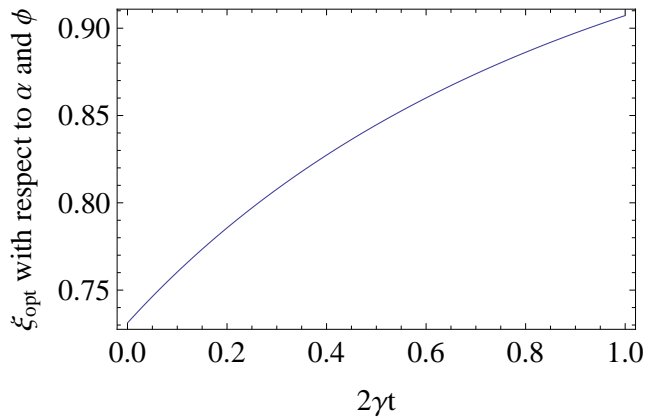}
   \caption{In the individual dephasing channel, the spin squeezing parameter of the weighted cluster state $\xi_2$
   as a function of $2\gamma t$ optimized with respect to $\alpha$ and $\phi$.}
   \label{figure5}
\end{figure}

\begin{figure}
   \includegraphics[width=8.5cm]{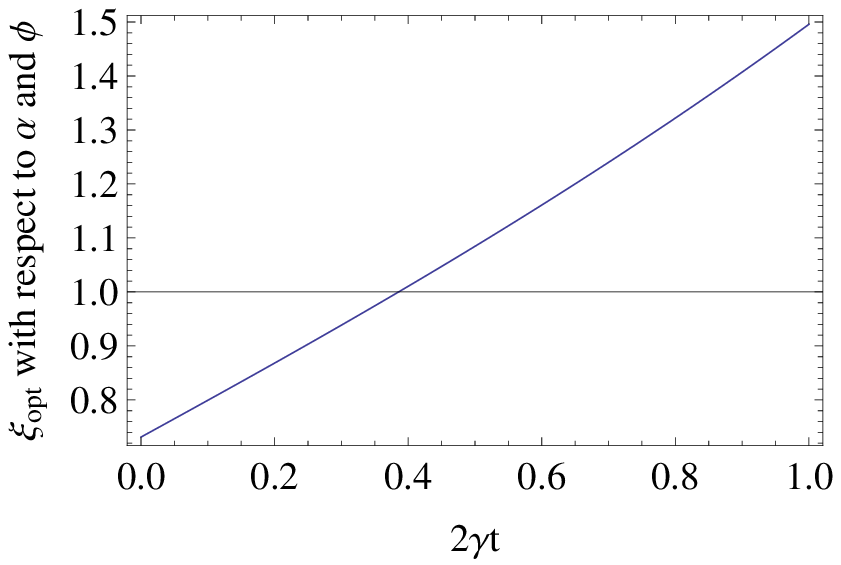}
   \caption{In the individual depolarizing channel, the spin squeezing parameter of the weighted cluster state $\xi_2$
   as a function of $2\gamma t$ optimized with respect to $\alpha$ and $\phi$.}
   \label{figure6}
\end{figure}

\begin{figure}
   \includegraphics[width=8.5cm]{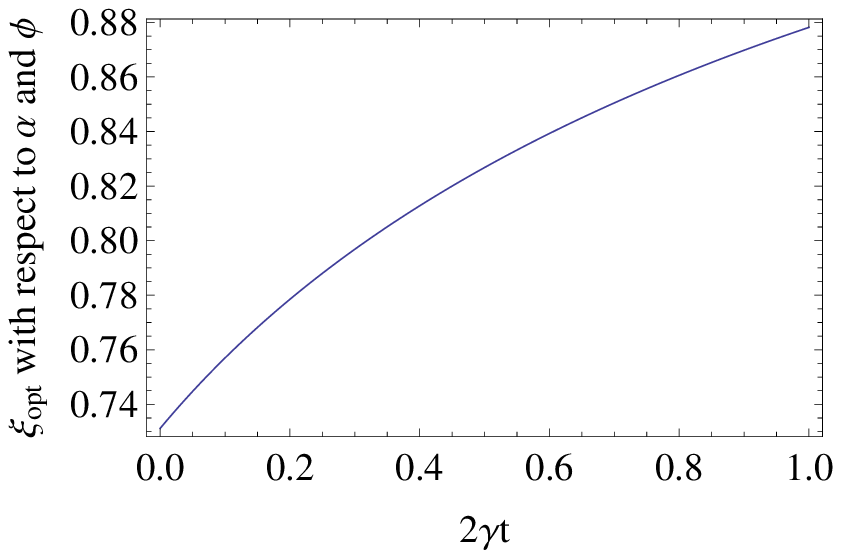}
   \caption{In the individual pure damping channel, the spin squeezing parameter of the weighted cluster state $\xi_2$
   as a function of $2\gamma t$ optimized with respect to $\alpha$ and $\phi$.}
   \label{figure4}
\end{figure}

For the weighted cluster state, we can easily evaluate the modified
main spin direction in the presence of dephasing as
\begin{equation}
\vec{n}'=\left(\cos\left(\Delta t-\alpha\right),\sin\left(\Delta
t-\alpha\right),0\right),
\end{equation}
and the modified orthogonal direction is
\begin{equation}
\vec{n}'_\perp=\left(-\cos\phi\sin\left(\Delta
t-\alpha\right),\cos\phi\cos\left(\Delta
t-\alpha\right),\sin\phi\right).
\end{equation}
Then the spin squeezing parameter of the weighted cluster state
evolves to
\begin{equation} \xi_2^2(t)=\frac{1+e^{-2\gamma
t}\cos^2\phi\sin^2\alpha/2+e^{-\gamma
t}\sin2\phi\sin\alpha}{\cos^4\alpha/2}.
\end{equation}
in the individual dephasing channel, which is the main type of
decoherence for a spin ensemble. Coupling to the individual
depolarizing channel, the spin squeezing parameter of the weighted
cluster state evolves to
\begin{equation}
\xi_2^2(t)=\frac{1+e^{-2\gamma t}\cos^2\phi\sin^2\alpha/2+e^{-\gamma
t}\sin2\phi\sin\alpha}{e^{-2\gamma t}\cos^4\alpha/2}.
\end{equation}
Whereas, the spin squeezing parameter of the weighted cluster state
evolves to
\begin{equation}
\xi_2^2(t)=\frac{1+e^{-\gamma t}\cos^2\phi\sin^2\alpha/2+e^{-\gamma
t}\sin2\phi\sin\alpha}{\left[1+e^{-\gamma
t}\left(\cos^2\alpha/2-1\right)^2\right]^2}
\end{equation}
in the individual damping channel.

Figs.~4-6 show the dynamics of the spin squeezing parameter of the
weighted cluster states coupled to different noisy channels. For
three different noisy channels, the optimized spin squeezing
parameter increases with time. In Figs.~4 and 6, coupled to the
individual dephasing and damping channel the spin squeezing property
of the weighted cluster state still holds ($<1$) for $0<2\gamma
t<1$. In Fig.~5, in the individual depolarizing channel, the spin
squeezing parameter of the weighted cluster state is smaller than
$1$ for small damping parameter $2\gamma t<0.398$.

For the weighted fully-connected graph state, we consider the same
dephasing channel. The modified mean spin direction and the
orthogonal direction are calculated as
\begin{align}
&\vec{n}'=\left(\cos\left(\Delta t-N\alpha\right),\sin\left(\Delta
t-N\alpha\right),0\right),\\
&\vec{n}'_\perp=\left(-\cos\phi\sin\left(\Delta
t-N\alpha\right),\cos\phi\cos\left(\Delta
t-N\alpha\right),\sin\phi\right).\nonumber
\end{align}
Coupling to the individual dephasing channel which is the main type
of decoherence for a spin ensemble, the spin squeezing parameter of
the weighted fully-connected graph state evolves to
\begin{equation}
\xi_2^2(t)=\frac{1+\frac{1}{4}e^{-2\gamma
t}\left(N-1\right)\left(A-\frac{A^2}{\sqrt{A^2+B^2}}\right)-\frac{1}{4}e^{-\gamma
t}\left(N-1\right)\frac{B^2}{\sqrt{A^2+B^2}}}{\cos^{2N-2}\alpha}.
\end{equation} Then the spin squeezing parameter of the weighted
fully-connected state evolves to
\begin{equation}
\xi_2^2(t)=\frac{1+\frac{1}{4}e^{-2\gamma
t}\left(N-1\right)\left(A-\frac{A^2}{\sqrt{A^2+B^2}}\right)-\frac{1}{4}e^{-\gamma
t}\left(N-1\right)\frac{B^2}{\sqrt{A^2+B^2}}}{e^{-2\gamma
t}\cos^{2N-2}\alpha}.
\end{equation}
under the individual depolarizing. Whereas, the spin squeezing
parameter of the weighted fully-connected graph state evolves to
\begin{equation}
\xi_2^2(t)=\frac{1+\frac{1}{4}e^{-\gamma
t}\left(N-1\right)\left(A-\frac{A^2}{\sqrt{A^2+B^2}}\right)-\frac{1}{4}e^{-\gamma
t}\left(N-1\right)\frac{B^2}{\sqrt{A^2+B^2}}}{\left[e^{-\gamma
t}\cos^{N-1}\alpha-\left(1-\text{e}^{-\gamma t}\right)\right]^2}.
\end{equation} in the individual damping channel.

\begin{figure}
   \includegraphics[width=8.5cm]{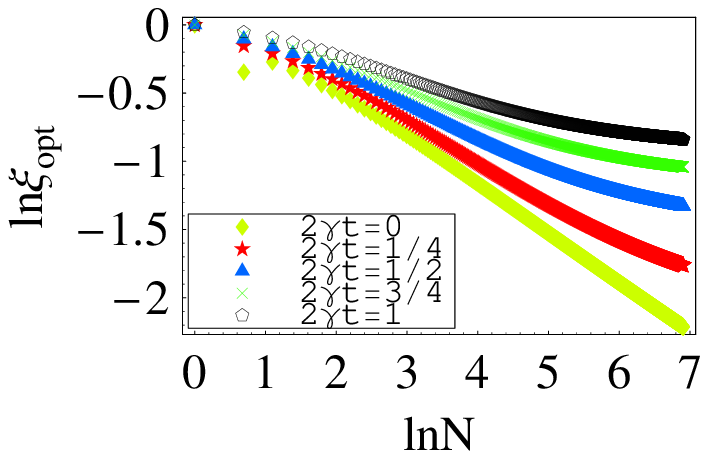}
   \caption{Ln-ln plots of the spin squeezing parameters $\xi_2$ of
   the weighted fully-connected graph state v.s. the number
   of the qubits $N$ (up to $10^3$) with different dephasing rate $\gamma$: $2\gamma t=0$ (yellow diamonds),
   $2\gamma t=0.25$ (red stars), $2\gamma t=0.5$ (blue triangles),
   $2\gamma t=0.75$ (green crosses), $2\gamma t=1$ (black circles).}
   \label{figure8}
\end{figure}

\begin{figure}
   \includegraphics[width=8.5cm]{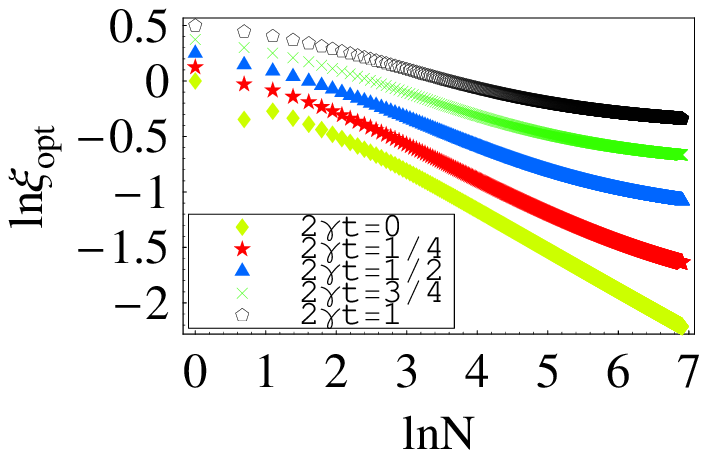}
   \caption{Ln-ln plots of the spin squeezing parameters $\xi_2$ of
   the weighted fully-connected graph state v.s. the number
   of the qubits $N$ (up to $10^3$) with different depolarizing rate $\gamma$: $2\gamma t=0$ (yellow diamonds),
   $2\gamma t=0.25$ (red stars), $2\gamma t=0.5$ (blue triangles),
   $2\gamma t=0.75$ (green crosses), $2\gamma t=1$ (black circles).}
   \label{figure9}
\end{figure}

\begin{figure}
   \includegraphics[width=8.5cm]{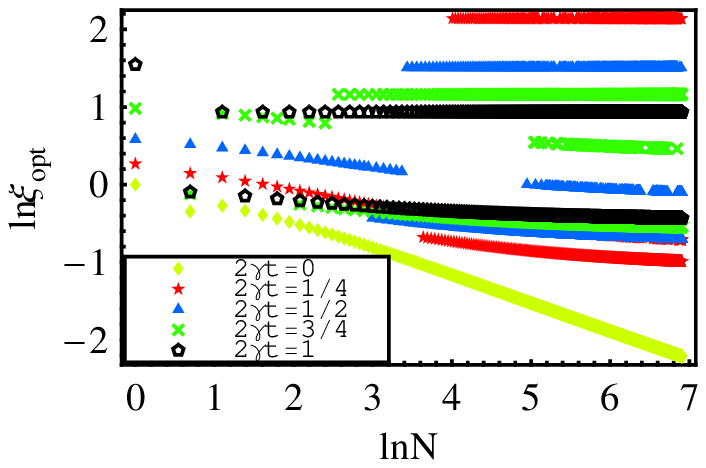}
   \caption{Ln-ln plots of the spin squeezing parameters $\xi_2$ of
   the weighted fully-connected graph state v.s. the number
   of the qubits $N$ (up to $10^3$) with different damping rate $\gamma$: $2\gamma t=0$ (yellow diamonds),
   $2\gamma t=0.25$ (red stars), $2\gamma t=0.5$ (blue triangles),
   $2\gamma t=0.75$ (green crosses), $2\gamma t=1$ (black circles).}
   \label{figure7}
\end{figure}

\begin{table}[t]
 \begin{tabular}{|c|c|c|c|c|c|} \hline
$ 2\gamma t$&$ \zeta $&$ \Delta \zeta $&$ \delta $\\
\hline $ 0 $&$ -0.3372 $&$ 0.009 $&$ 0.2730 $\\
\hline
$ 0.25 $&$ -0.2643$&$ 0.013 $&$ 0.0298 $\\
\hline $
0.5 $&$ -0.1813 $&$ 0.014 $&$ -0.1173 $\\
\hline $ 0.75 $&$ -0.1348 $&$ 0.022 $&$ -0.1455 $\\
\hline $ 1 $&$ -0.1064 $&$ 0.030 $&$ -0.1338 $\\
\hline
 \end{tabular}
 \caption{The linear regression data
    $\text{ln} \xi_2=(\zeta \pm \Delta \zeta)\text{ln} N+\delta$ of the dynamics of the spin squeezing of
    weighted fully-connected graph state in the individual dephasing channel with different dephasing parameters.}
 \label{table1}
\end{table}

\begin{table}[t]
 \begin{tabular}{|c|c|c|c|c|c|} \hline
$ 2\gamma t$&$ \zeta $&$ \Delta \zeta $&$ \delta $\\
\hline $ 0 $&$ -0.3372 $&$ 0.006 $&$ 0.2730 $\\
\hline
$ 0.25 $&$ -0.2643$&$ 0.009 $&$ 0.1548 $\\
\hline $
0.5 $&$ -0.1813 $&$ 0.013 $&$ 0.1327$\\
\hline $ 0.75 $&$ -0.1348 $&$ 0.030 $&$ 0.2295 $\\
\hline $ 1 $&$ -0.1064 $&$ 0.034 $&$ 0.3662 $\\
\hline
 \end{tabular}
 \caption{The linear regression data
    $\text{ln} \xi_2=(\zeta \pm \Delta \zeta)\text{ln} N+\delta$ of the dynamics of the spin squeezing of
    weighted fully-connected graph state in the individual depolarizing channel with different depolarizing parameters.}
 \label{table2}
\end{table}

Figs.~7-9 show ln-ln plots of the spin squeezing parameter $\xi_2$
of the weighted fully-connected graph state coupled to different
noisy channels including dephasing, depolarizing and damping
channels v.s. the number of qubits $N$ optimized with respect to
$\alpha$. In Figs.~7 and 8 the simulated results on the spin
squeezing parameter of the weighted fully-connected graph state
coupled to the individual dephasing and depolarizing channels
respectively are shown and both follow a power law on the particle
number $N$, thus $\text{ln}\xi_2=\zeta\text{ln} N+\delta$.
Corresponding linear regression data presented in Tables I and II
clear reveals slopes decrease with the decoherence rates increasing.
Fig. 9 shows the spin squeezing of weighted fully-connected graph
state coupled to individual damping channel, which is not linear any
more but for large $N$ the spin squeezing property still holds.

\section{Applications}

As we mentioned in Sec.~II, the spin squeezing can be used to
improve the frequency standards. In this section we show that spin
squeezing property of weighted fully-connected graph states are
robust to practical decoherence. In the absence of decoherence,
using the product state with $N$ qubits the phase sensitivity is
$\delta\varphi\propto1/\sqrt{N}$ and using the $N$-qubit maximally
entangled state such as GHZ-type state the phase sensitivity can be
improved to be $\delta \varphi\propto1/N$. Hence, the squeezed
uncertainty of a spin component directly improves the sensitivity in
interferometry including Ramsey spectroscopy, enabling the
improvement factors up to $\sqrt{N}$ for GHZ-type states with $N$
particles. In the presence of decoherence, one will obtain the
similar accuracy for frequency standards for GHZ-type states
as~\cite{HM97}
\begin{equation}
\delta\varphi_\text{GHZ}=\frac{\sqrt{2\gamma t
e}}{\sqrt{N}}\propto\frac{1}{\sqrt{N}}.
\end{equation} That means there is no
improvement even with GHZ-type maximally entangled states. However
with one of partial entangled states---weighted fully-connected
graph states the sensitivity is still reduced in the presence of
decoherence but our scheme remains superior over several reference
schemes with states such as GHZ-type states and product states
because of the robustness of spin squeezing property.

In linear interferometry, the phase precision for an $N$-particle
weighted fully-connected graph state can be explained by classical
statistics. The situation is equivalent to $N$ individual
measurements on a single particle~\cite{G+10}. The minimal phase
error for a classical measurement $\delta \varphi=1/\sqrt{N}$ is
known as the standard quantum limit. In the case of correlated
particles, the classical limit can be exceeded. The idea is to
achieve improved scaling of the interferometric phase sensitivity
with the number of particles using entangled states within the
interferometer~\cite{GLM04,PS09} as low as the Heisenberg limit,
where the phase error is $\delta \varphi=1/N$. A weighted
fully-connected graph state is a special kind of many-particle spin
squeezing state for which the minimal interferometric phase error is
\begin{equation}
\delta \varphi=\frac{\xi_2}{\sqrt{N}}.
\end{equation}
If the relation between the phase error $\delta\phi$ with the
particle number is a power law, then we define
\begin{equation}
\text{ln}\delta\varphi=-\varsigma \text{ln}
N+\text{ln}\delta\varphi_0,
\end{equation}
with $\varsigma=1$ for the Heisenberg limit, $\varsigma=1/2$ for the
classical limit and $1/2<\varsigma<1$ means that the weighted
fully-connected graph state can still be used to improve the
frequency standard.

\begin{figure}
   \includegraphics[width=8.5cm]{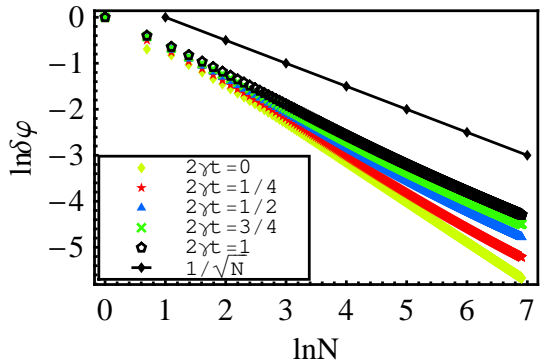}
   \caption{Ln-ln plots of the phase error $\delta \varphi$ v.s. the number
   of the qubits $N$ (up to $10^3$) with different dephasing rate $\gamma$: $2\gamma t=0$ (yellow diamonds),
   $2\gamma t=0.25$ (red stars), $2\gamma t=0.5$ (blue triangles),
   $2\gamma t=0.75$ (green crosses), $2\gamma t=1$ (black circles). The solid black line
   presents the shot-noise limit $\delta \varphi\propto 1/\sqrt{N}$.}
   \label{figure11}
\end{figure}

\begin{figure}
   \includegraphics[width=8.5cm]{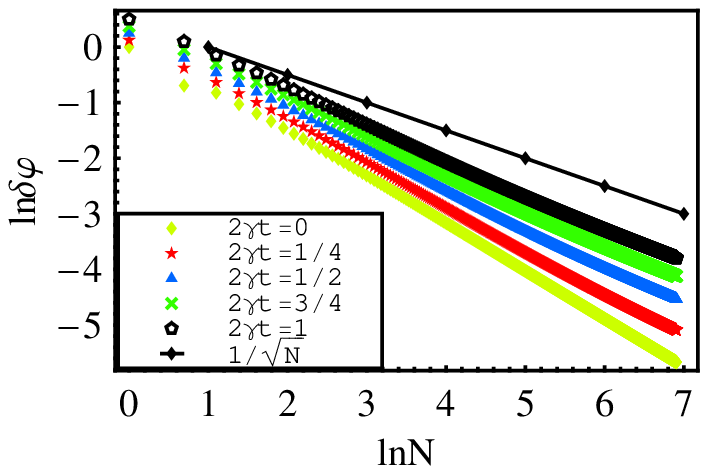}
   \caption{Ln-ln plots of the phase error $\delta \varphi$ v.s. the number
   of the qubits $N$ (up to $10^3$) with different depolarizing rate $\gamma$: $2\gamma t=0$ (yellow diamonds),
   $2\gamma t=0.25$ (red stars), $2\gamma t=0.5$ (blue triangles),
   $2\gamma t=0.75$ (green crosses), $2\gamma t=1$ (black circles). The solid black line
   presents the shot-noise limit $\delta \varphi\propto 1/\sqrt{N}$.}
   \label{figure12}
\end{figure}

\begin{figure}
   \includegraphics[width=8.5cm]{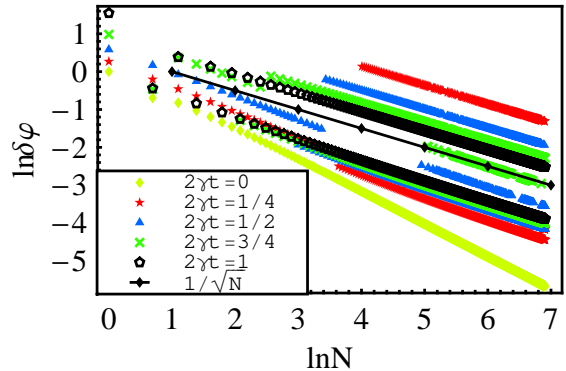}
   \caption{Ln-ln plots of the phase error $\delta \varphi$ v.s. the number
   of the qubits $N$ (up to $10^3$) with different damping rate $\gamma$: $2\gamma t=0$ (yellow diamonds),
   $2\gamma t=0.25$ (red stars), $2\gamma t=0.5$ (blue triangles),
   $2\gamma t=0.75$ (green crosses), $2\gamma t=1$ (black circles). The solid black line
   presents the shot-noise limit $\delta \varphi\propto 1/\sqrt{N}$.}
   \label{figure10}
\end{figure}

In the presence of decoherence, Huelge {\it et al.}~\cite{HM97} have
shown that a GHZ-type maximally entangled state does not provide an
improved sensitivity compared to product states. We show here that a
weighted fully-connected graph state with a high symmetry gives an
improved sensitivity under the three kinds of
decoherence---dephasing, depolarizing and damping.

For example, by optimizing with respect to the weights $\alpha$, the
spin squeezing parameter of the weighted fully-connected graph state
is obtained numerically $\xi_2\propto1/N^{0.1064}$ with the
decoherence rate (dephaisng or depolarizing) satisfying $2\gamma
t=1$ shown in Figs.~7 and 8. Thus the phase sensitivity $\delta
\varphi\propto1/N^{0.6064}$ follows a power law on the number of the
particles not a constant in Figs.~10 and 11. In Fig.~12 just as the
spin squeezing parameter, ln-ln plot of the phase sensitivity under
the damping v.s. the particle number is not linear any more. However
from the slopes of the plot we can see find for some $N$ the spin
squeezing holds and the sensitivity can still be improved with the
weighted fully-connected graph state coupled to individual damping
channels.

We compare the schemes on improve the phase sensitivity with
weighted fully-connected graph states and product states and prove
that with weighted fully-connected graph states an improvement in
the phase sensitivity is achieved beyond shot-noise limit. Now we
compare our scheme to that with GHZ-type entangled states with same
decoherent rate and evolution time. We define an improvement in the
sensitivity
\begin{equation}
P=\frac{\delta\varphi_\text{GHZ}}{\delta\varphi}=\frac{\sqrt{2\gamma
t e}}{\xi_2}.
\end{equation}
If $P>1$ the phase error in the scheme with a weighted
fully-connected graph states is smaller than that with a GHZ state
with same decoherent rate and evolution time and we have an
improvement in the sensitivity. Remarkably, Figs.~13-15 show that
for small $N$, $P<1$ that means the phase error in the scheme with
GHZ states is smaller than that with weighted full-connected graph
states. However with $N$ increasing our scheme shows its advantage.
From numerical calculations with $N$ up to $10^3$, we have an
improvement $P\sim 4.5$ under dephasing and $P\sim 3.5$ under
depolarizing with respect to the shot noise limit $\sim 1/N^{0.5}$.

\begin{figure}
   \includegraphics[width=8.5cm]{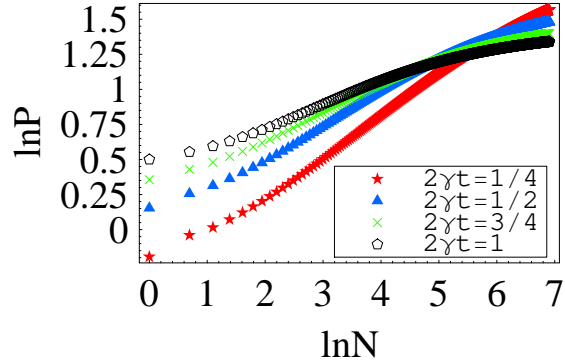}
   \caption{Ln-ln plots of the improvement in the sensitivity $P$ v.s. the number
   of the qubits $N$ (up to $10^3$) with different dephasing rate $\gamma$: $2\gamma t=0$ (yellow diamonds),
   $2\gamma t=0.25$ (red stars), $2\gamma t=0.5$ (blue triangles),
   $2\gamma t=0.75$ (green crosses), $2\gamma t=1$ (black circles).}
   \label{figure13}
\end{figure}

\begin{figure}
   \includegraphics[width=8.5cm]{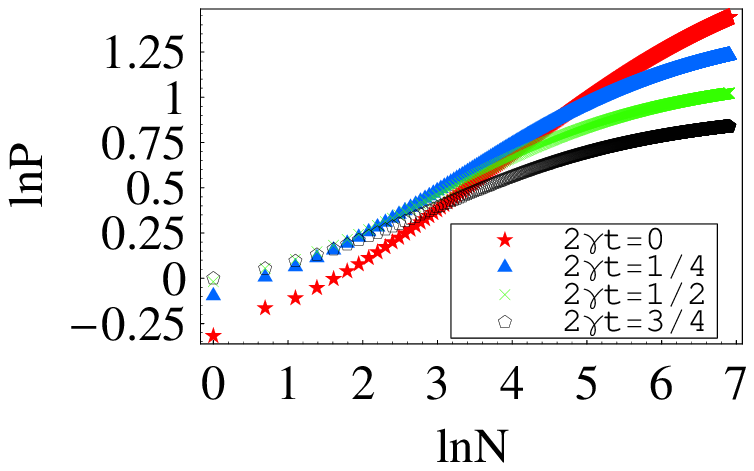}
   \caption{Ln-ln plots of the improvement in the sensitivity $P$ v.s. the number
   of the qubits $N$ (up to $10^3$) with different depolarizing rate $\gamma$: $2\gamma t=0$ (yellow diamonds),
   $2\gamma t=0.25$ (red stars), $2\gamma t=0.5$ (blue triangles),
   $2\gamma t=0.75$ (green crosses), $2\gamma t=1$ (black circles).}
   \label{figure14}
\end{figure}

\begin{figure}
   \includegraphics[width=8.5cm]{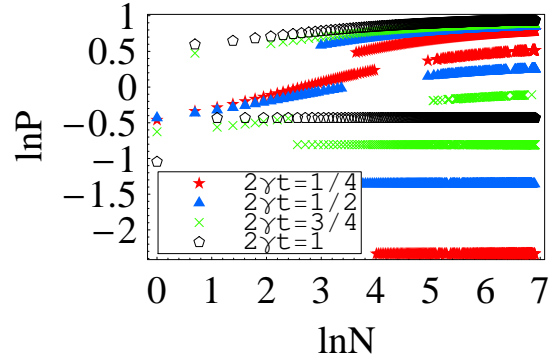}
   \caption{Ln-ln plots of the improvement in the sensitivity $P$ v.s. the number
   of the qubits $N$ (up to $10^3$) with different damping rate $\gamma$: $2\gamma t=0$ (yellow diamonds),
   $2\gamma t=0.25$ (red stars), $2\gamma t=0.5$ (blue triangles),
   $2\gamma t=0.75$ (green crosses), $2\gamma t=1$ (black circles).}
   \label{figure15}
\end{figure}

\section{Conclusion}

The Heisenberg scaling $1/N$ in the decoherence-free case can be
achieved. In the absence of decoherence, both of the weighted graph
states and maximally entangled states are used to improve
interferometric sensitivity. Whereas, in the presence of decoherence
the ability of improving sensitivity is due to the robustness
against decoherence. The open question is if the decoherence limit
in the improvement without any measurement optimization is achieved
with all the known weighted graph states. Our goal is to prove that
it is spin squeezing instead of entanglement which improves the
sensitivity and find out the weighted graph states whose spin
squeezing property is robust to decoherence. Based on the analysis,
in the presence of decoherence, when $N$ is large enough, the phase
sensitivity can not be improved any more via the maximally entangled
states and product states , while the scheme with weighted graph
states still survive and $\delta\varphi$ beats the shot-noise limit
without any measurement optimization. Furthermore, one can obtain
the optimal improvement of sensitivity by tuning the weighted of
each edges and choosing proper joint measurements.

\begin{acknowledgments}
This work was performed during a sabbatical year at Key Lab of
Quantum Information, Chinese Academy of Sciences. The author thanks
Yongsheng Zhang and Xiangfa Zhou for useful conversations. This work
has been supported by the National Natural Science Foundation of
China under Grant Nos 11004029 and 11174052, the Natural Science
Foundation of Jiangsu Province under Grant No BK2010422, the Ph.D.\
Program of the Ministry of Education of China, the Excellent Young
Teachers Program of Southeast University and the National Basic
Research Development Program of China (973 Program) under Grant No
2011CB921203.
\end{acknowledgments}

\end{document}